\documentclass[onecolumn,amsmath,amssymb]{qm2008_abs}
\usepackage{graphicx}
\usepackage{dcolumn}
\usepackage{bm}
\newcommand{ \be }{\begin{eqnarray}}
\newcommand{ \ee }{\end{eqnarray}}
\newcommand{ \ben }{\begin{enumerate}}
\newcommand{ \een }{\end{enumerate}}
\newcommand{ \la }{\langle}
\newcommand{ \ra }{\rangle}

\def\P{$\cal P$}
\def\CP{$\cal CP$}
\newcommand{ \psirp }{\Psi_{RP}}

\newcommand{\mean}[1]{\left< {#1} \right>}

\usepackage{color}

\begin{document}
\title
{\Large Probe for the 
strong
parity violation effects  
at RHIC
with three particle correlations
}
\bigskip
\bigskip
\author{\large Sergei A. Voloshin for the STAR Collaboration}

\affiliation{
Wayne State University, Michigan 48201, USA }

\email{voloshin@wayne.edu}
\bigskip
\bigskip

\begin{abstract}
In non-central relativistic heavy ion collisions, \P-odd domains,
which might be created in the process of the collision,
are predicted to lead to charge separation
along the system orbital momentum~\cite{Kharzeev:2004ey}. 
An observable, \P-even, but directly sensitive to the charge
separation effect, has been proposed in~\cite{Voloshin:2004vk} 
and is based on 3-particle mixed harmonics azimuthal correlations.
We report the STAR measurements using this observable for Au+Au and
Cu+Cu collisions at $\sqrt{s_{NN}}$=200 and 62~GeV.
The results are reported as function of collision centrality, particle
separation in rapidity, and particle transverse momentum.
Effects that are not related to parity violation
but might contribute to the signal are discussed.
\end{abstract}

\maketitle

\section{Introduction. Effect and observables.}

It was suggested in~\cite{Kharzeev:1998kz} that metastable 
\P~and/or \CP-odd domains,                 
characterized by non-zero topological charge, might be created in 
ultra-relativistic heavy ion collisions. The possibility for
experimental detection of this phenomena have been discussed 
in~\cite{Voloshin:2000xf,Finch:2001hs}. 
More recently, it was noticed~\cite{Kharzeev:2004ey} that    
in non-central collisions such domains 
can demonstrate itself via preferential same charge particle
emission along the system angular momentum.
Such charge separation is a consequence of a difference in the number
of particles with positive and negative helicities positioned
in a strong magnetic field of  
non-central nuclear collision~\cite{Kharzeev:2004ey,Kharzeev:2007jp}.
Depending on the sign of the domain's topological charge, positively charged
particles will be preferentially emitted either along, or in the opposite
direction of the  system orbital momentum, with negative particles 
flowing oppositely to the positives. 
For a particular domain the charge separation 
can be effectively described by azimuthal distribution
\be
 dN_\pm /d\phi \propto (1+2 a_\pm \sin(\phi-\psirp)),
\label{eq:a}
\ee
where parameter $a_- = -a_+$, and the sign of $a_\pm$ varies event
to event following the fluctuations in the domain's topological
charge. On average, $\mean{a_\pm}=0$, therefore one has to 
 measure $\mean{a_\alpha a_\beta}$. 
The latter is a \P-even
quantity and may contain contribution from 
effects not related to the parity violation. 
A correlator, directly sensitive to $\mean{a_\alpha a_\beta}$
was proposed in~\cite{Voloshin:2004vk}:
\be
\hspace*{-2cm}
\la \cos(\phi_\alpha +\phi_\beta -2\psirp) \ra 
\label{eq:obs1}
&=&
\mean{\cos(\phi_\alpha  -\psirp)\cos(\phi_b  -\psirp)} 
\\
\nonumber
&&
-\mean{\sin(\phi_\alpha  -\psirp)\sin(\phi_b  -\psirp)}
\approx
(v_{1,a}v_{1,b} - a_\alpha a_b), 
\ee
and represents the
difference in correlations projected onto the reaction plane
and ``out-of-plane'' direction~\cite{Borghini:2001vi,Adams:2003zg}.
The approximate sign in the last equality reflects the fact that
particles $\alpha$ and $\beta$ might be correlated not only via common
correlation to the reaction plane -- then the average does not factorize
into product of averages. 
Note that 
in 
the rapidity region symmetric with respect to the mid-rapidity, the average
directed flow equals to zero. 
In practice one estimates the (second order) reaction plane with the
third particle~\cite{Poskanzer:1998yz,Borghini:2001vi,Adams:2003zg}
assuming that the particle $c$ is correlated with particles
$\alpha$ and $\beta$ only via common correlation to the reaction plane: 
$
\la \cos(\phi_a +\phi_\beta -2\phi_c) \ra 
=
\la \cos(\phi_a +\phi_\beta -2\psirp) \ra \, v_{2,c}. 
$  

\section{Data. Detector.}

The data have been obtained with STAR detector during RHIC
Runs IV and V of Au+Au and Cu+Cu collisions at $\sqrt{s_{NN}}=200$
and 62~GeV. Minimum bias trigger has been used and events with primary
vertex within 30~cm from the center of the Main TPC have been
selected. Standard STAR cuts suppressing pile-up event have been applied.  
The results are  based on about 14.7M Au+Au and 13.9M Cu+Cu events 
at  $\sqrt{s_{NN}}=200$~GeV, and  2.4M Au+Au and  6.3M Cu+Cu events at
 $\sqrt{s_{NN}}=62$~GeV. The centrality of the collision is determined
according to the charged particle multiplicity in the region $|\eta|<0.5$.

The correlations are reported for pseudorapidity 
region $|\eta|<1.0$ covered by the STAR Main TPC. 
For the event plane determination, 
in addition to the main TPC we use two Forward
TPCs with pseudorapidity coverage $2.7<|\eta|<3.9$, and two ZDC-SMD
(Zero Degree Calorimeter - Shower Maximum Detector)
detectors that being sensitive to directed flow of neutrons in
the beam rapidity region.
The tracks in the main TPC are required to have $p_t>0.15$~GeV/c. 
For the results integrated over transverse momentum we also impose upper
cut of $p_t<2$~GeV/c.
Standard STAR track quality cuts are applied: the minimum of 15
tracking points are required for a track and the ratio of number of points 
to the maximum possible is required 
to be greater than 0.52 to avoid the effects of
track splitting. The data with reverse magnetic field
polarity has been used to asses the systematic effects; the reported
results are averaged over both field polarities.

\section{Results and Discussion}

\begin{figure}[ht]
  \includegraphics[width=.47\textwidth]{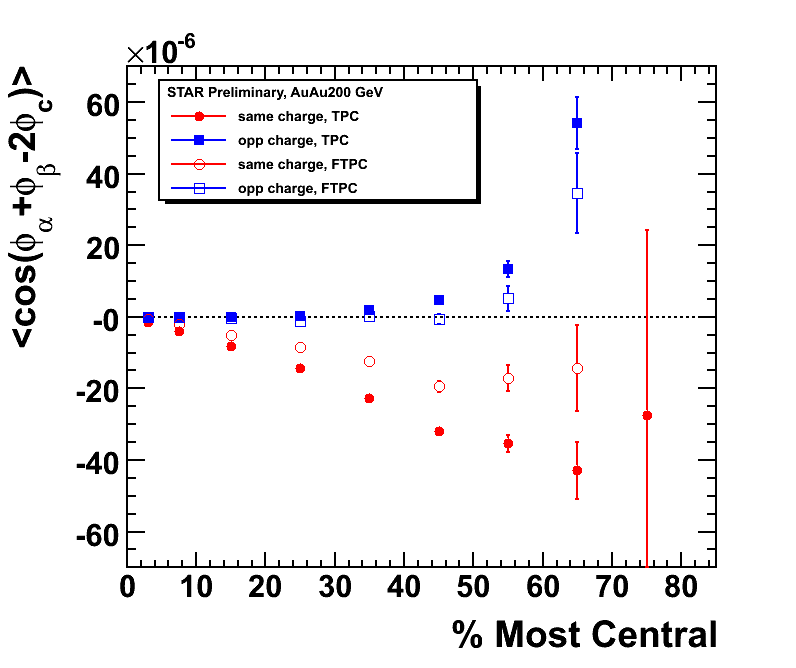}
  \includegraphics[width=.47\textwidth]{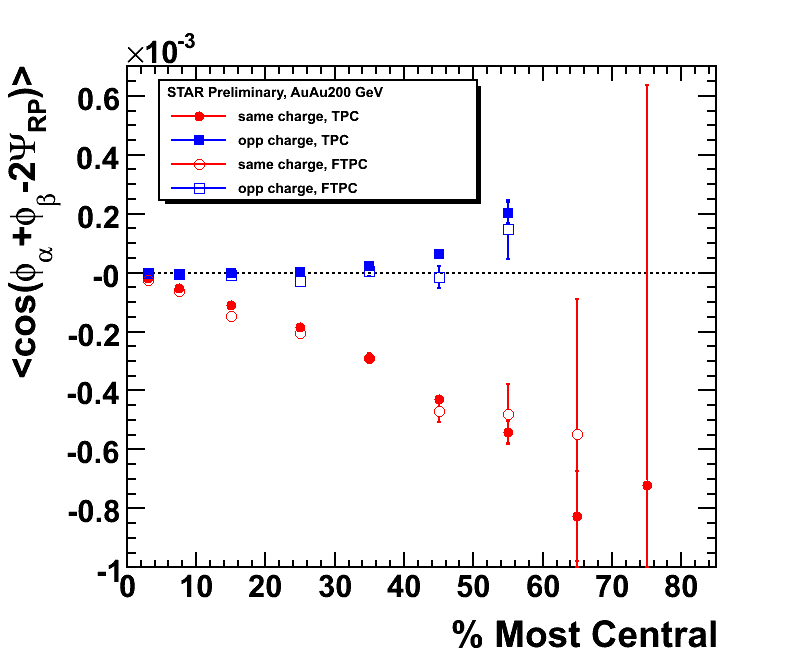}
\centerline{(a) \hspace{.47\textwidth} (b) }
  \caption{(a) Comparison of correlations obtained using third particle in
  the main TPC and Forward TPCs. (b) The results
  after normalization to the flow of the third particle.}
  \label{fig:au200F}
\end{figure}

\begin{figure}[ht]
  \includegraphics[width=.47\textwidth]{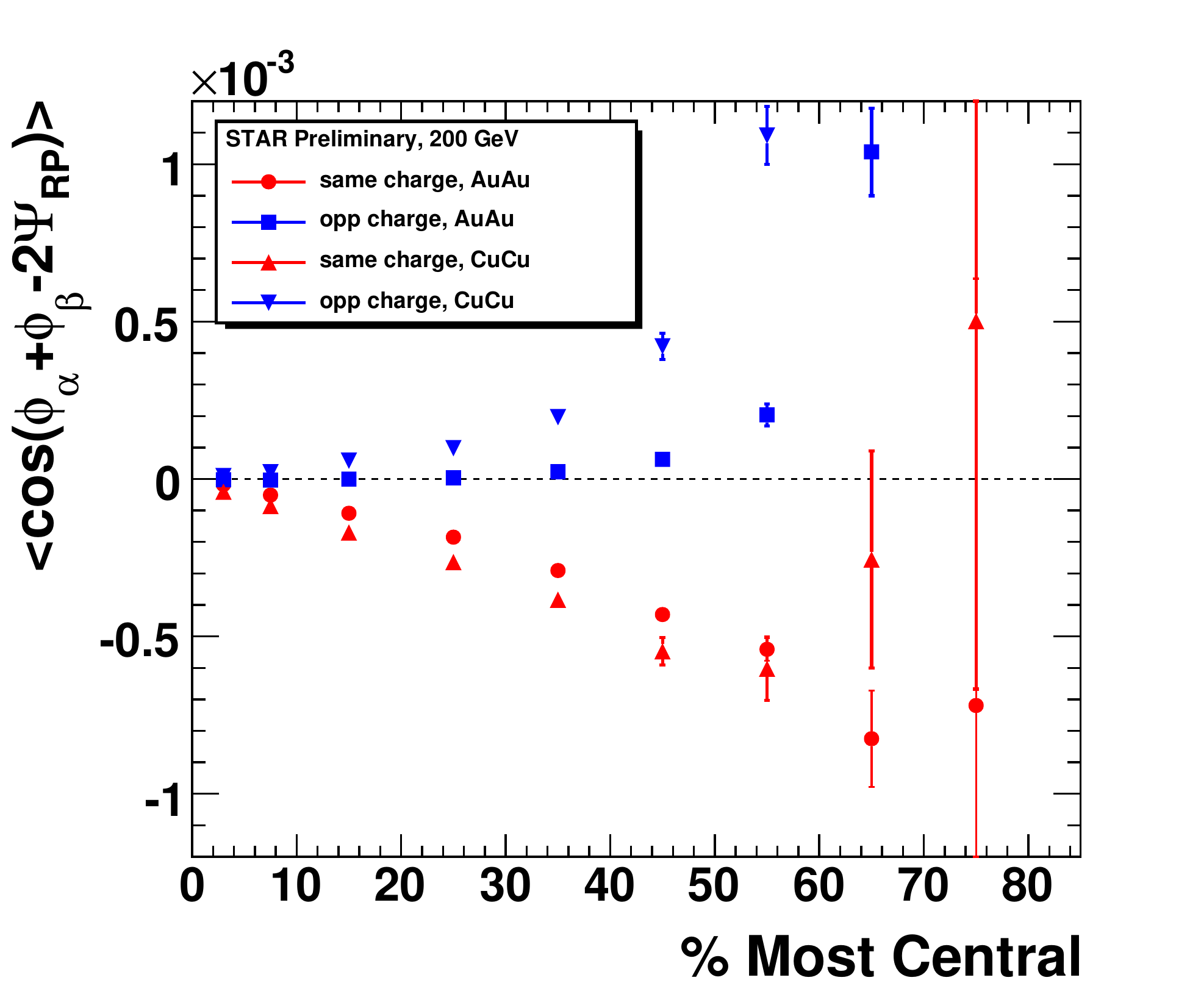}
  \includegraphics[width=.47\textwidth]{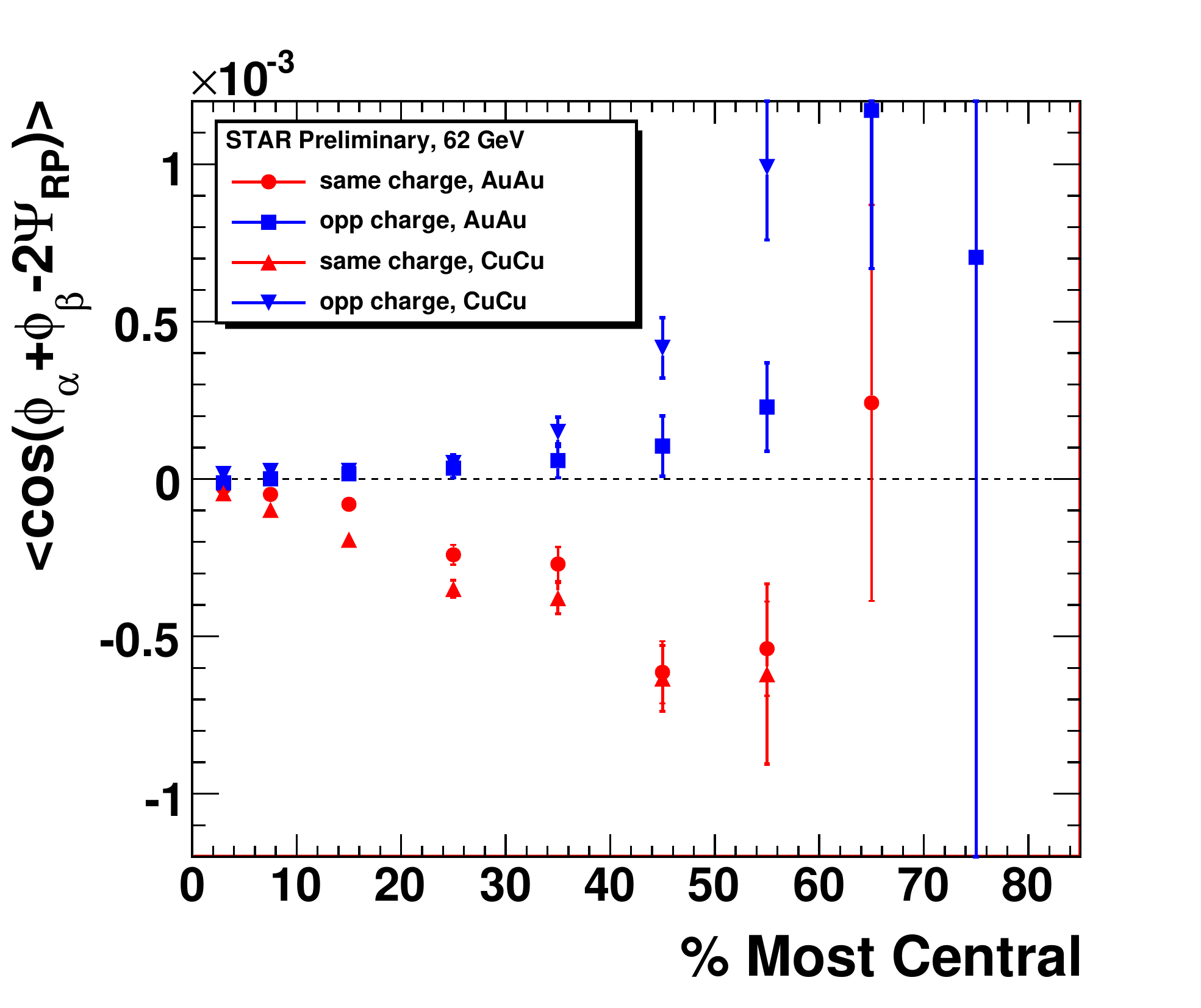}
\centerline{(a) \hspace{.47\textwidth} (b) }
  \caption{ $\mean{\cos(\phi_a +\phi_\beta -2\psirp) }$
in Au+Au and Cu+Cu
collisions at (a) $\sqrt{s_{NN}}=200$~GeV and (b) 62~GeV. 
}
  \label{fig:uuv2}
\end{figure}

\begin{figure}[ht]
  \includegraphics[width=.44\textwidth]{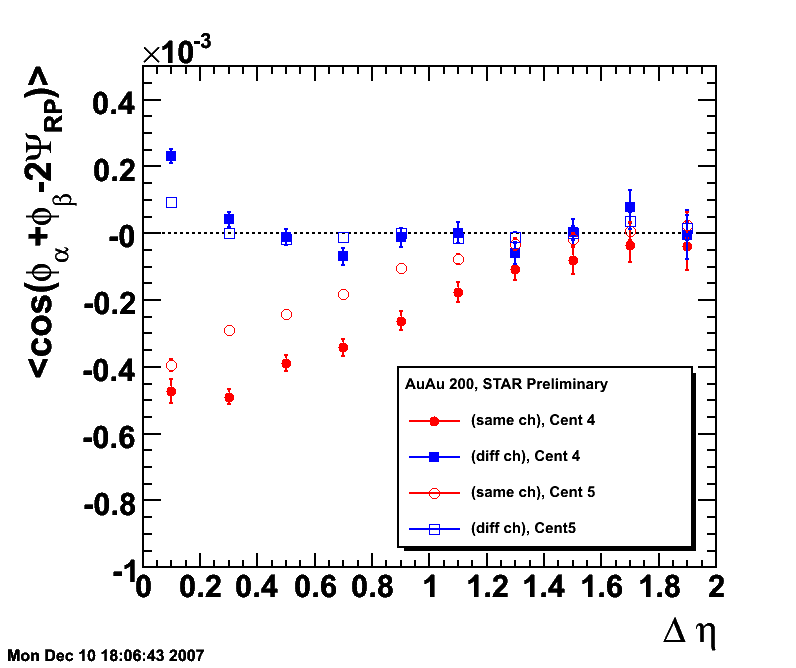} 
  \includegraphics[width=.47\textwidth]{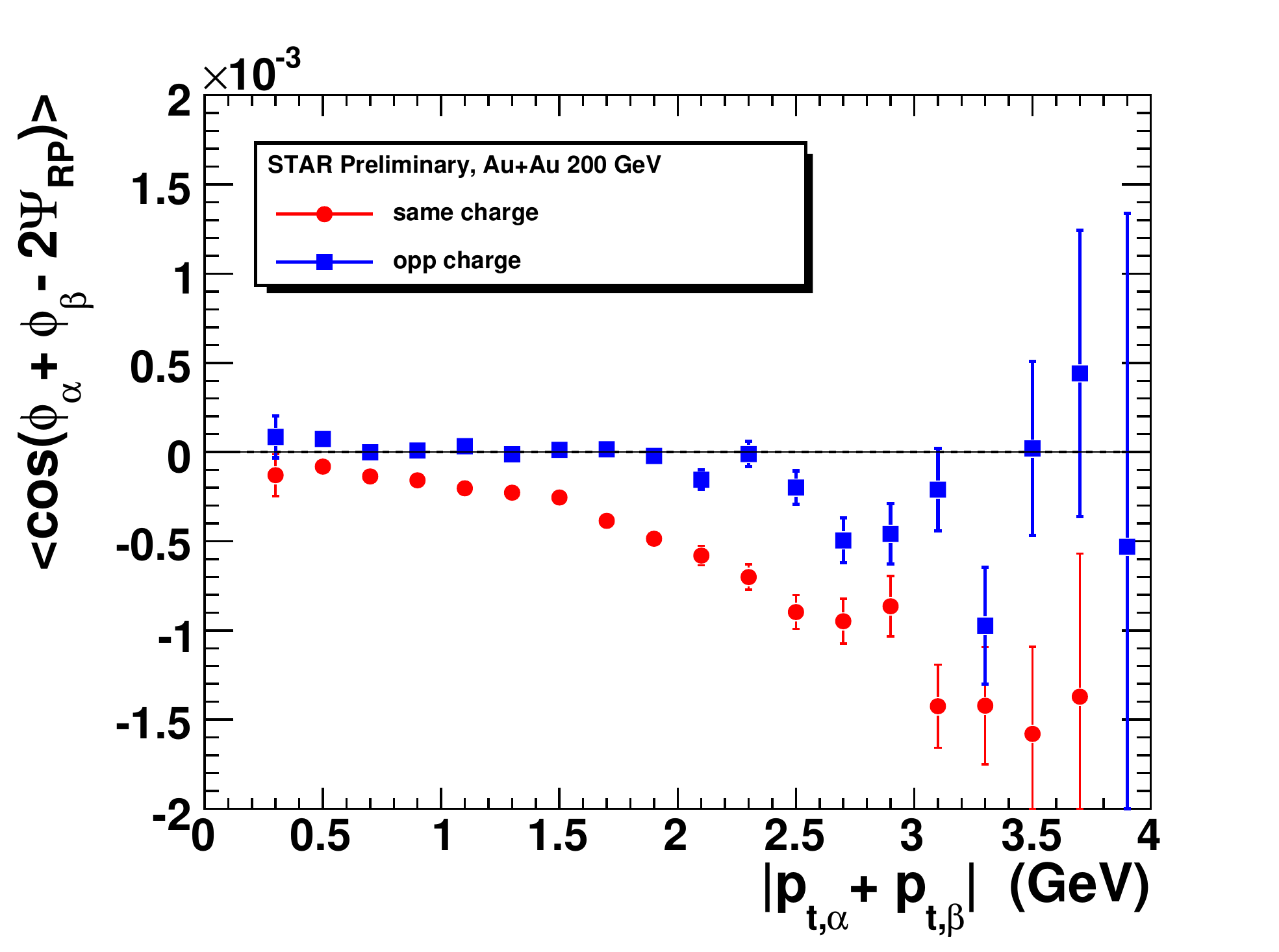}
\centerline{(a) \hspace{.47\textwidth} (b) }
  \caption{Au+Au 200 GeV. The signal dependence on
(a) $|\eta_\alpha -\eta_\beta|$ and 
(b) $p_{t,\alpha}+p_{t,\beta}$.} 
  \label{fig:uuv2diff}
\end{figure}

Fig.~\ref{fig:au200F}(a) compares three particle correlations 
obtained for different charge combinations
as function of centrality of the collision with the third particle
selected in the Main TPC or Forward TPC regions. 
Under assumption that the third (second harmonic) 
particle is correlated with the first (first harmonic) two particles
only via common correlation to the reaction plane, this correlator
should be proportional to the elliptic flow of the third particle.
As seen from Fig.~\ref{fig:au200F}(b), which  shows three
particle correlator divided by $v_2$ of the third particle, our
results agree well with this assumption. 
Similar conclusion can be also reached if ZDC-SMD is used for the
determination of the reaction plane. In the following all results are
presented normalized by flow of the particles used in the
determination of the event plane.
Fig.~\ref{fig:uuv2} presents the results for Au+Au and Cu+Cu
collisions at two different energies. The signal in Cu+Cu collisions
seems to be somewhat larger for the same centrality of the collision.
The opposite sign combination
correlations are stronger in Cu+Cu, qualitatively in agreement with
scenario of stronger suppression of the back-to-back correlations in
Au+Au collisions.
In this and other figures, only statistical errors are indicated. 
Comparing the results obtained with two different magnetic field
polarities we observe a systematic difference in the results 
comparable to the signal in the 5\% most central collisions;
negligible for other centralities. 
Other systematic uncertainties are currently under study.
Fig.~\ref{fig:uuv2diff}(a) shows the dependence of the signal on the
difference in pseudorapidities of two particles for two centralities.
The signal has a ``typical hadronic'' width of about 1 unit of
pseudorapidity.
Fig.~\ref{fig:uuv2diff}(b) shows the dependence of the signal on the
sum the magnitudes of two particles for one
centrality.
We do not observe that the signal is concentrated in the low $p_t$
region as expected for \P-violation effects. At the same time our $p_t$
range is 
statistics limited and further
study is needed for better determination of the dependence of the
signal on transverse momentum.  

\section{Other effects}

The correlator, Eq.~\ref{eq:obs1} 
can exhibit non-zero
signal for effects not related to \P-violation, 
e.g. 
when particles $\alpha$ and $\beta$ are
products of a cluster 
decay, and the cluster itself exhibits
elliptic flow~\cite{Borghini:2001vi,Adams:2003zg} or decay/fragment differently
when emitted in-plane or out-of-plane.
One can write:
\be
\la \cos(\phi_\alpha + \phi_\beta -2\phi_c) \ra 
\approx \frac{f_{res} \, \la  \cos(\phi_\alpha + \phi_\beta -2\phi_{res}) \ra
\; v_{2,res}}{N_{ch}} \, v_{2,c},
\ee
where $f_{res}$ is the fraction of charged particles 
originating from corresponding resonance
decays, 
$v_{2,res}$ is the resonance elliptic flow. 
The factor $1/N_{ch}$ 
reflects the probability  that both particles in the pair are
from the same resonance.  
Note that  $ \la  \cos(\phi_\alpha + \phi_\beta -2\phi_{res}) \ra $ is zero if
the resonance is at rest, and become non-zero only due to resonance
motion.

We have 
investigated 
possible contributions of not \P-violation effects
known to us: so far we could not identify any that would explain even
qualitatively the observed signal.
(i) The contribution of {\em directed flow fluctuations} is of opposite
sign and is similar for different charge combinations unlike the
signal.
(ii) 
The contribution of {\em  (elliptically) flowing hadronic
resonances} has been found too small in magnitude, 
somewhat stronger for opposite
sign pairs, and of the same sign for all charge combinations.
(iii) 
The effect of strong (elliptically modulated) radial flow 
has been studyied with
PYTHIA even generator. The signal, though in magnitude of the same
order as in the data,  has been found similar for 
different charge combinations unlike in the data.
(iv) Global polatization of hyperons has been found experimentally to
be consistent with zero~\cite{Abelev:2007zk}.

\section{Summary}

The analysis using three particle correlations that are directly
sensitive to the \P-violation effects in heavy ion collisions has
been presented for Au+Au and Cu+Cu collisions at $\sqrt{s_{NN}}$=200
and 62~GeV.
The results are reported for different particle charge combinations 
as function of collision centrality, particle
separation in rapidity, and particle transverse momentum.
Qualitatively the results agree with the magnitude and gross features
of the theoretical predictions for \P-violation in heavy ion
collisions, except, probably, transverse momentum dependence.
Though a particular observable used in our analysis is
\P-even and in principle might be sensitive to other, not parity
violating, effects, so far, with systematics checks mentioned above, 
we could not identify such that would explain the
observed correlations.


\end{document}